# Near Field Scanning Optical Imaging of Gold Nanoparticles in the Sub-Wavelength Limit

Light-matter Interaction of metallic Au Nanoparticles

Prajit Dhara
Department of Electrical and Electronics Engineering,
Birla Institute of Technology and Science Pilani
prajitdhara1@gmail.com

A.K.Sivadasan
Nanomaterials and Sensors Section, Indira Gandhi Centre for Atomic Research, Homi Bhabha National Institute
sivankondazhy@gmail.com

*Abstract*— **The near-field scanning optical microscopic (NSOM) imaging of Au nanoparticles with size in the sub-wavelength limit (<λ/2N.A.) is reported. The NSOM imaging technique can resolve the objects which is beyond the scope of optical microscope using visible light (λ ≈ 500 nm) with objectives having a numerical aperture (N.A.) close to unity. The role of evanescent waves which is an exponentially decaying field with higher momenta *i.e.*, lower wavelengths compared to that of normal light, in the metal dielectric interface is realized for imaging of noble metal nanostructures with sub-wavelength dimension in the near field. However, the confined light with components of evanescent waves, emanating from the NSOM probe, interacts with the oscillating dipoles present in the sub-diffraction limited nanostructures and produce propagating waves, which can be recorded by the far field detector. The light-matter interactions of Au nanoparticles of diameters in the range of 10-150 nm probed by the NSOM technique with a visible excitation of 532 nm are reported. The strong surface plasmon resonance (SPR) related absorption of Au nanoparticles is envisaged for explaining the contrast variations in the recorded NSOM images.**

## I. INTRODUCTION

The study of optical properties of nanostructures in dimensions below the half wavelength of light is a very interesting and challenging task. Abbe's diffraction limit prevents conventional optical microscopes to possess a spatial resolution beyond the value of ~λ/2 (sub-wavelength limit), where λ is the wavelength of excitation with a maximum numerical aperture value of unity for the probing objective [1]. Therefore even visible light of λ=400 nm cannot image nanostructures of size below 200 nm.

The imaging techniques such as scanning electron microscope (SEM) and atomic force microscope (AFM) are limited to produce only the information about the dimensions of nanostructures. Presently, the integration of light with electronic circuits is a hot research area in the scientific community [2]. The nanostructures are very useful in this aspect for coupling of light to the sub-wavelength regime and hence the miniaturization of photonic and optoelectronic devices. Therefore, the light-matter interactions in the sub-diffraction limited nanostructures are very important in terms of fundamental research as well as its applications in the various industries [3].

The light-matter interactions in metallic nanostructures have opened to a new branch of surface plasmon (SP) based photonics, known as plasmonics [4]. The near field scanning optical microscope (NSOM) assisted with the help of plasmonics is a tool well equipped to visualize simultaneously the topographic information as well as the light-matter interaction in the near field regime of nanostructures. The light passing through the metal coated tip of NSOM probe with a circular aperture of diameter around few nanometers at the apex is capable of surpassing the diffraction limit due to surface plasmonic effects. In the near field regime, the emitted evanescent field from the NSOM probe is not diffraction limited. Hence, it facilitates optical and spectroscopic imaging of objects with spatial resolution upto a few nanometers. The SPs originate due to the collective oscillation of the free electrons about the fixed positive charge centers in the surface of metal nanostructures with a frequency of quantized harmonic oscillation of electrons, also known as plasma frequency ($\omega_p$) expressed by

$$\omega_p = \left(\frac{n_e e^2}{m_{eff}\, \varepsilon_0}\right)^{1/2} \ldots\ldots\ldots\ldots\ldots\ldots\ldots\ldots\ldots(1)$$

where, $n_e$ is the density, $m_{eff}$ is the effective mass, $e$ is the charge of an electron and $\varepsilon_0$ is the permittivity of free space [1,3,5]. The coupling of the incident electromagnetic waves with the coherent oscillation of free-electron plasma near the metal surface is known as a surface plasmon polariton (SPP) and it is a propagating surface wave at the continuous metal-dielectric interface. The electromagnetic field perpendicular to the metal surface decays exponentially and is hence known as evanescent wave, providing sub-wavelength confinement near to the metal surface. Matching of the incident excitation frequency ($\omega$) of electromagnetic wave with the plasmon frequency ($\omega_p$) of the electrons in metal nanostructures, leads to an enhanced light-matter interaction, known as surface plasmon resonance (SPR).

In the present report, we have utilized NSOM as a technique for imaging and characterization of metallic Au nanoparticles (diameter ~50-150 nm) with an external laser excitation of 532 nm (≈2.33 eV).

## II. EXPERIMENTAL SECTION

The Au thin film was coated on a Si (100) substrate for 5 min in a thermal evaporation chamber (12A4D, HINDHIVAC, India) with a constant evaporation rate (1Å /sec). In order to make uniform sized Au nanoparticles, the as-grown Au film was annealed at a temperature 900 °C for 10 min in an inert (Ar) atmosphere. The synthesized Au nanoparticles were studied using the multi function scanning probe microscopy (SPM) system. The AFM-images as well as NSOM images were recorded by using 150 nm tip configured with a tuning fork feedback mechanism in the four probe SPM coupled with Raman spectroscopy (MultiView 4000; Nanonics, Israel).

An optical fiber with a circular aperture and metal (Au/Cr) coated probe with a tip apex (aperture) diameter of 150 nm, was used for the near field excitation. The metallic coating of Cr and Au on the fiber is used to avoid leakage of optical power, enhanced optical transmission and ensured confinement of the light to the sample surface. The scanning was performed either by using the NSOM probe or motorized XY sample stage with very precise spatial resolution controlled by inbuilt sensors and piezo-drivers. Excitation of 532 nm was used as it corresponds to the reported SPR of Au nanoparticles in the range of 500-550 nm [6]. A band pass filter (532 nm ≈ 2.33 eV) was used to extract the excitation laser after the light-matter interaction and before reaching the light to the detector in the far field configuration. The detailed schematic representation of the experimental setup for NSOM is available in one of our earlier report [7]. The same probe was used as an AFM tip for simultaneous scanning of the topography along with the NSOM image of the sample with the tuning fork feedback mechanism. The tuning fork feedback mechanism was adapted in this study instead of the usual beam bounce feedback mechanism as the experiment involved the use of a different light source for NSOM.

## III. RESULTS AND DISCUSSION

The low magnification AFM topographic images of Au nanoparticles on the Si (100) substrate are shown in the figure 1. The study was carried out using an NSOM probe with a tip size of 150 nm. The two dimensional (2D) image (Fig. 1(a)) shows smooth, spherical shape in morphology, uneven size and well-dispersed distribution of Au nanoparticles. The 3D (Fig. 1(b)) AFM topographic image also shows the similar features as in 2D image along with variation in height distribution of Au nanoparticles. Since the diameter of Au nanoparticles (~50-150 nm) is far below the diffraction limit for the excitation wavelength (532 nm), one needs to shorten the wavelength down to the sub-diffraction regime to obtain highly resolved optical images.

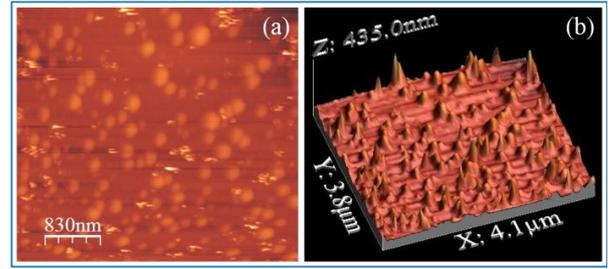

Fig.1 The AFM topographic images of Au nanoparticles on the Si (100) substrate in (a) 2D and (b) 3D.

Using metal coated NSOM probe, it is possible to produce evanescent waves with momentum higher than that of the original excitation wavelength $\lambda_0=2\pi/k_0(\omega)$ with wave vector of $k_0(\omega)=\omega/c$, where $c$ is the velocity of light. Therefore, the confined light waves with evanescent components emanating from the NSOM probe aperture possess group of wave vectors higher than the original excitation laser as $k_{ev}(\omega)=\omega/v$, with different velocities ($v$) slower than the excitation wave velocity ($v<c$) [5]. Therefore, the NSOM measurements are advantageous to provide super-resolution along with the localization of intense electric fields. At the same time, it conserve the excitation energy and hence the frequency. Thus, it offers the possibility of optical as well as spectroscopic imaging in the sub-wavelength regime. So, from the scanned images using NSOM technique, we can understand even the intrinsic properties of a sample as revealed by its electronic or vibrational characteristics of the material.

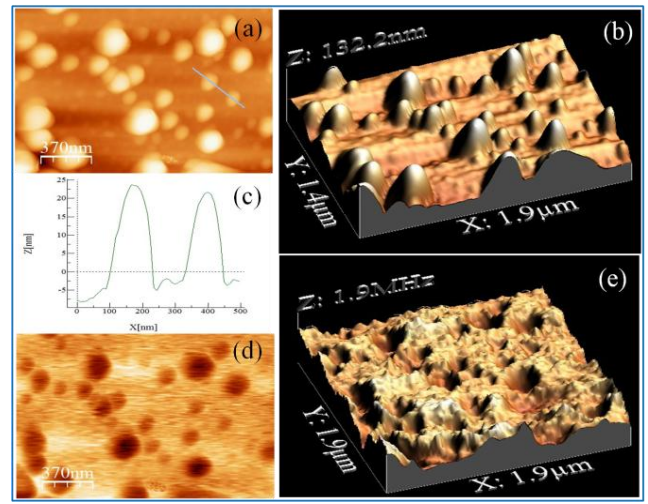

Fig. 2 AFM topographic images of Au nanoparticles in a selected area of the sample with 150 nm tip (a) 2D and (b) 3D. (c) The height profile of Au nanoparticles along the line drawn in the figure 2(a). The corresponding NSOM images of the sample collected in the same area in (d) 2D and (e) 3D

The investigation of near field interaction for metallic Au nanoparticles was performed using a visible laser light of energy 2.33 eV (~532 nm). The results obtained from the NSOM imaging of the predefined area of the sample using a 150 nm sized probe are as shown in the figure 2. The AFM topographic images of Au nanoparticles in a selected area of the sample are shown in 2D (Fig. 2(a)) and 3D ((Fig. 2(b)). The height profile of Au nanoparticles along the line drown in the figure 2(a) is shown in the figure 2(c). It is very clear that the diameter of the nanoparticles is ~30 nm. The NSOM image of Au nanoparticles recorded from the same area of the sample (Figs. 2(d) and 2(e)) shows a strong absorption of electromagnetic waves. The significant absorption of light with wavelength 532 nm by Au nanoparticle is because of the fact that, the generation of localized SPR (LSPR) of light with plasmons. The peak value of ~535 nm in the broad absorption spectrum (Fig. 3) for Au nanoparticles (inset Fig. 3) matches with value of the excitation wavelength which clearly indicates that the absorption observed in the NSOM images are due the presence of LSPR.

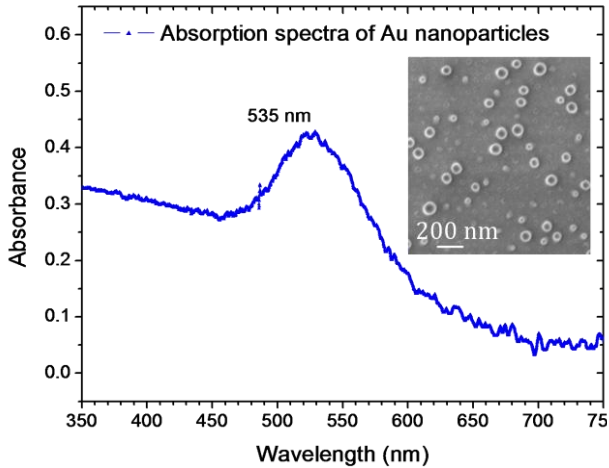

Fig. 3 Absorbance plot of typical Au NPs. Inset figure shows the FESEM image of Au nanoparticles

At resonance, the incident electromagnetic waves can also couple with collective oscillation of electrons, can produce SPPs which are perpendicular to the surface of the Au nanoparticles. The frequency dependent wave vector of SPP can be expressed in terms of frequency dependent dielectric constants of metal ($\varepsilon_m = \varepsilon_m' + i\varepsilon_m''$) and surrounding dielectric material ($\varepsilon_d = 1$, for air or vacuum), as

$$k_{spp}(\omega) = \frac{\omega}{c}\sqrt{\frac{\varepsilon_d \cdot \varepsilon_m}{\varepsilon_d + \varepsilon_m}}$$

Therefore, the effective wavelength of the SPP is $\lambda_{spp} = 2\pi/k_{spp}$ [1,3,5]. The SPPs of different wavelengths, lower than the excitation, can propagate through the surface of Au nanoparticles up to a propagation length which depends on the complex dielectric constants of the metal and dielectric medium [1,3,5]. Once the SPP propagates through the surface of Au nanoparticle and crosses the metallic region, then the electromagnetic wave may decouple from the SPP and it can be converted to a propagating wave. The intensity of the absorption is influenced by the frequency dependent poalrizability of the Au nanoparticles and it can be varied with respect to the size of the Au nanoparticles. Thus, because of the variation of different sizes of the Au nanoparticles, it is possible to observe them with relatively different absorption intensities (Figs. 2(d) and 2(e)). Apart from the formation of SPP, some portion of the absorbed excitation laser may also participate in the lattice phonon generations leading to heating as well as inter-band transitions of Au NPs [8].

IV. CONCLUSIONS

In conclusion, we use the NSOM technique for optical imaging of Au nanostructures of sub-wavelength dimension in the near field regime. The NSOM images of metallic Au nanoparticles with diameters in the range of 10-150 nm are imaged using 150 nm NSOM probe and 532 nm excitation. The NSOM images of metallic Au nanoparticles shows a strong surface plasmon resonance related absorption of excitation laser with wavelength of 532 nm due to the localized surface plasmon resonance of Au nanoparticles ~535 nm as well as the formation of surface plasmon polaritons.

ACKNOWLEDGMENT

One of us (PD) would like to thank Avinash Patsha, Kishore K. Madapu and Raktima Basu of Nanomaterials and Sensors Section, Surface Nanoscience Division, Materials Science Group, IGCAR who put in as much effort as himself for this work.